\documentclass[usenatbib]{mn2e}
\usepackage{amsfonts}
\usepackage{amsmath}
\usepackage{amssymb}
\usepackage{graphicx}
\usepackage{color}

\def\lsim{~\rlap{$<$}{\lower 1.0ex\hbox{$\sim$}}}
\def\bsim{~\rlap{$>$}{\lower 1.0ex\hbox{$\sim$}}}

\def\hmpc{\ {\rm {\it h}^{-1}Mpc}}

\def\hmsun{\ {\rm M_\odot/{\it h}}}

\def\la{\langle}
\def\ra{\rangle}

\def\ln{{\rm ln}}

\def\vk{\mathrm{\bf k}}

\def\vw{\mathrm{\bf w}}
\def\vx{\mathrm{\bf x}}
\def\vy{\mathrm{\bf y}}

\def\npk{n_{\rm pk}}
\def\nesp{n_{\rm ESP}}

\def\bnesp{\bar{n}_{\rm ESP}}
\def\dpk{\delta_{\rm pk}}

\title[Measuring nonlocal Lagrangian peak bias]
      {Measuring nonlocal Lagrangian peak bias}

\author[M. Biagetti, K.C. Chan, V. Desjacques \& A. Paranjape]
{Matteo Biagetti$^{1}$, Kwan Chuen Chan$^{1}$, 
Vincent Desjacques$^{1}$\thanks{E-mail: Vincent.Desjacques@unige.ch} \& Aseem Paranjape$^{2}$ \\  
 $^1$ D\'epartement de Physique Th\'eorique and 
      Center for Astroparticle Physics (CAP), Universit\'e de Gen\`eve, \\ 
      24 quai Ernest Ansermet, CH-1211 Gen\`eve, Switzerland \\
 $^2$ Instt. for Astronomy, Dept. of Physics, ETH Z\"urich, Wolfgang-Pauli-Strasse 27, CH-8093 Z\"urich, Switzerland}

\newcommand{\fesp}{f_{\rm ESP}}

\definecolor{RoyalBlue}{rgb}{0.25,.41,.88}

\begin{document}
\pagerange{\pageref{firstpage}--\pageref{lastpage}}

\maketitle 

\label{firstpage}

\begin{abstract}
We investigate nonlocal Lagrangian bias contributions involving gradients 
of the linear density field, for which we have predictions from the excursion 
set peak formalism. We begin by writing down a bias expansion which includes all 
the bias terms, including the nonlocal ones. Having checked that the model 
furnishes a reasonable fit to the halo mass function, we develop a 1-point 
cross-correlation technique to measure bias factors associated with 
$\chi^2$-distributed quantities.
We validate the method with numerical realizations of peaks of Gaussian random 
fields before we apply it to N-body simulations. We focus on the lowest (quadratic) 
order nonlocal contributions $-2\chi_{10}(\vk_1\cdot\vk_2)$ and 
$\chi_{01}[3(\vk_1\cdot\vk_2)^2-k_1^2 k_2^2]$, where $\vk_1$, $\vk_2$ are wave
modes.
We can reproduce our measurement of $\chi_{10}$ if we allow for an offset 
between the Lagrangian halo center-of-mass and the peak position. The sign and 
magnitude of $\chi_{10}$ is consistent with Lagrangian haloes sitting near 
linear density maxima. The resulting contribution to the halo bias can safely be 
ignored for $M=10^{13}\hmsun$, but could become relevant at larger halo masses. 
For the second nonlocal bias $\chi_{01}$ however, we measure a much larger 
magnitude than predicted by our model. We speculate that some of this discrepancy 
might originate from nonlocal Lagrangian contributions induced by nonspherical 
collapse.
\end{abstract}


\section{Introduction}
\label{sec:intro}

Understanding the clustering of dark matter haloes has been a topic of active 
research for many years. A number of analytic approaches have been developed 
to tackle this issue such as the peak model \citep{bardeen/bond/etal:1986},
the excursion set framework \citep{bond/cole/etal:1991} or perturbation theory 
\cite[see e.g.][for a review]{bernardeau/colombi/etal:2002}.
Heuristic arguments like the peak-background split \citep{kaiser:1984}, and
approximations like local bias \citep{fry/gaztanaga:1993} have been very 
helpful for modelling the clustering of dark matter haloes. 
Nevertheless, improvements in computational power and numerical algorithms as 
well as the advent of large scale galaxy surveys have considerably increased 
the need for an accurate description of halo clustering. Until recently however, 
it was unclear how the peak approach, which is thus far the only framework in 
which biased tracers form a discrete point set, relates to the more widespread 
excursion set theory, local bias approximation or peak-background split argument. 

Working out this connection has been the subject of several recent papers.
\cite{desjacques:2013}, building on earlier work by \cite{desjacques/crocce/etal:2010}, 
showed that correlation functions of discrete density peaks can be computed 
using an effective (i.e. which does not involve measurable counts-in-cells 
quantities) generalized bias expansion in which all the bias parameters, 
including those of the nonlocal terms~\footnote{To facilitate the comparison 
with other studies, we will call nonlocal terms all contributions to Lagrangian 
clustering that are not of the form $\delta^n(\vx)$, where $\delta(\vx)$ is the 
linear mass density field.}, can be computed from a peak-background split. 
In parallel, \cite{paranjape/sheth:2012} demonstrated how the peak formalism, 
which deals with statistics of density maxima at a fixed smoothing scale, can
be combined with excursion set theory, whose basic building block is the 
density contrast at various filtering scales. 
Similar ideas can already be found in the early work of \cite{bond:1989}.
\cite{paranjape/sheth/desjacques:2013} (hereafter PSD) subsequently computed 
the mass function and linear bias of haloes within this excursion set peak 
(ESP) approach and showed that it agrees very well with simulation data. 

The focus of this work is on the second-order nonlocal bias terms 
predicted by the ESP approach. These generate corrections to the Fourier 
peak bias of the form $-2\chi_{10}(\vk_1\cdot\vk_2)$ and 
$\chi_{01}[3(\vk_1\cdot\vk_2)^2-k_1^2 k_2^2]$ \citep{desjacques:2013}. What 
makes them quite interesting is the fact that there are related to $\chi^2$ 
rather than normally-distributed variables. Here, we will show how one can 
measure their amplitude in the bias of dark matter haloes without computing 
any correlation function. Of course, this technique can also be applied to
measure nonlocal Lagrangian bias contributions induced by e.g. the tidal 
shear, but this will be the subject of future work.

This paper is organized as follows. In a first part, we will advocate 
a slight modification of the original excursion set peak formulation of PSD 
in order to easily write down the corresponding effective bias expansion 
(Sec. \S\ref{sec:ESP}). 
Next, we will explain how the cross-correlation technique proposed by 
\cite{musso/paranjape/sheth:2012}, which has already been successfully applied 
to the bias factors associated with the density field
\citep{paranjape/sheth/desjacques:2013,paranjape/sefusatti/etal:2013},
can be extended to measure the second-order nonlocal bias factors $\chi_{10}$ 
and $\chi_{01}$ that weight the two quadratic, nonlocal bias contributions 
(Sec. \S\ref{sec:CHICROSS}).
Finally, we will validate our method with peaks of Gaussian random fields 
before measuring $\chi_{10}$ and $\chi_{01}$ for dark matter haloes 
(Sec. \S\ref{sec:SIMS}). We conclude in \S\ref{sec:conclusion}.

\section{Excursion set peaks}
\label{sec:ESP}

In this section we apply the excursion set approach to the peak model in the case 
of a moving barrier to get a prediction of the halo mass function which we compare 
to simulations. We then get expressions for bias parameters, generalising results 
in \cite{desjacques:2013,desjacques/gong/riotto:2013}. We also point out a few 
changes to PSD.
We show that, as far as the mass function is concerned, these modifications do not 
make much difference (only few percents, in agreement with what PSD found), but 
they affect first- and second-order bias parameters, as new terms arise.

\subsection{Notation}

We will adopt the following notation for the variance of the smoothed density field
(linearly extrapolated to present-day) and its derivatives, 
\begin{equation}
\sigma^2_{j\alpha} = \frac{1}{2\pi^2}\int_0^\infty\!\! dk\, P(k) 
k^{2(j+1)} W_\alpha^2(kR_\alpha) \;,
\end{equation}
where $P(k)$ is the power spectrum of the mass density field, $W_\alpha(kR_\alpha)$ 
and the subscript $\alpha=G$ or $T$ will denote Gaussian or tophat filtering, 
respectively. Moreover, $R_\alpha$ is the Lagrangian smoothing scale (which may 
depend on the choice of kernel).
Denoting $\delta_T$ and $\delta_G$ the linear density field smoothed with a tophat
and Gaussian filter, respectively, we introduce the variables
\begin{align}
\nu(\vx) &= \frac{1}{\sigma_{0T}}\delta_T(\vx) \\ 
u(\vx) &=-\frac{1}{\sigma_{2G}}\nabla^2\delta_G(\vx) \nonumber \\ 
\mu(\vx) &= -\frac{d\delta_T}{dR_T}(\vx) \nonumber \;.\label{eq:mu}
\end{align}
Note that, while $\nu$ and $u$ have unit variance, $\mu$ is not normalized. We will 
use the notation $\langle \mu^2 \rangle = \Delta^2_0$ in what follows.

Cross-correlations among these three variables are useful and will be denoted as
\begin{align}
\langle \nu u \rangle &= \gamma_1 = \frac{\sigma_{1X}^2}{\sigma_{0T}\sigma_{2G}} \\
\langle \nu \mu \rangle &= \gamma_{\nu\mu} = 
\frac{1}{\sigma_{0T}}\int_0^\infty\!\! \frac{dk}{2\pi^2}\,P(k) 
k^2 W_T(kR_T)\frac{dW_T(kR_T)}{dR_T} \\
\langle u\mu \rangle &= \gamma_{u\mu} =  
\frac{1}{\sigma_{2G}}\int_0^\infty\!\! \frac{dk}{2\pi^2}\, P(k) 
k^4 W_G(kR_G)\frac{dW_T(kR_T)}{dR_T} \label{eq:mu}\;.
\end{align}
The first-order, mixed spectral moment $\sigma_{1X}$ is
\begin{equation}
\label{eq:sigma1x}
\sigma^2_{1X} = \frac{1}{2\pi^2}\int dk P(k) k^4 W_T(kR_T)W_G(kR_G) \;,
\end{equation}
i.e. one filter is tophat and the other Gaussian.

\subsection{First-crossing and moving barrier}

\subsubsection{Summary of previous results}

Let us first summarize the basic ideas behind the excursion set peaks approach 
introduced by \cite{paranjape/sheth:2012} and further developed in PSD and 
\cite{desjacques/gong/riotto:2013}.

The excursion set approach states that a region of mass $M$ has virialized when the
overdensity $\delta(R)$, where $R\sim M^{1/3}$ is the filtering scale associated with 
the perturbation, reaches the spherical collapse threshold $\delta_c$ provided that, 
for any $R'>R$, the inequality $\delta(R)<\delta_c$ holds. 
This last condition formally implies an infinite set of contraints (one at each 
smoothing scale). However, as was shown in \cite{musso/sheth:2012}, the requirement 
$\delta(R+\Delta R)<\delta_c$ with $\Delta R \ll 1$ furnishes a very good approximation. 
This follows from the fact that the trajectory described by $\delta(R)$ as a function 
of $R$ is highly correlated for large radii. As a result, if $\delta$ crosses $\delta_c$ 
at $R$, then it is almost certainly below the threshold at any larger radius.

This first-crossing condition can be combined with the peak constraint, so that peaks 
on a given smoothing scale are counted only if the inequality above is satisfied. In
this case, the effective peak bias expansion introduced in \cite{desjacques:2013} is
modified through the presence of a new variable $\mu$ (Eq.\ref{eq:mu}) which, as was 
shown in \cite{desjacques/gong/riotto:2013}, reflects the dependence of bias to the 
first-crossing condition.

\subsubsection{Modifications to \cite{paranjape/sheth/desjacques:2013}}

We made a couple of modifications to the approach of PSD, which we will now describe
in more details.

Firstly, PSD used the fact that $\mu\equiv u$ when Gaussian filtering is also applied to 
the density field, so that the first-crossing condition can be accounted for with the 
variable $u$ only. When $\delta$ is smoothed with a tophat filter however, one should 
in principle deal explicitly with $\mu$ and, therefore, consider the trivariate normal 
distribution ${\cal N}(\nu,u,\mu)$. We will proceed this way.

Secondly, \cite{sheth/mo/tormen:2001} argued that, owing to the triaxiality of collapse, 
the critical density for collapse is not constant and equal to $\delta_c=1.68$, but  
rather distributed around a mean value which increases with decreasing halo mass.
Analyses of N-body simulations have confirmed this prediction and showed the scatter 
around the mean barrier is always significant 
\citep{dalal/white/etal:2008,robertson/kravtsov/etal:2009,Elia:2011ds}.
Since the stochasticity induced by triaxial collapse is somewhat cumbersome to implement 
in analytic models of halo collapse \citep[see e.g.][for a tentative implementation with
the peak constraint]{hahn/paranjape:2013}, we will 
consider a simple approximation calibrated with numerical simulations \citep[note that it 
differs from the diffusing barrier approach of][]{maggiore/riotto:2010}.
Namely, the square-root stochastic barrier 
\begin{equation}
\label{eq:barrier}
B= \delta_c + \beta \sigma_0 \;,
\end{equation}
wherein the stochastic variable $\beta$ closely follows a lognormal distribution, 
furnishes a good description of the critical collapse threshold as a function of halo 
mass \citep{robertson/kravtsov/etal:2009}. 
In PSD, this result was interpreted as follows: each halo ``sees'' a moving barrier 
$B=\delta_c+\beta\sigma_0$ with a value of $\beta$ drawn from a lognormal distribution.
Therefore, the first-crossing condition becomes
\begin{equation}
B < \delta < B + \left(B' + \mu\right) \Delta R \;,
\end{equation}
where the prime designates a derivative w.r.t. the filtering scale. Here however, we will
assume that each halo ``sees''  a constant (flat) barrier, whose height varies from halo
to halo. Therefore, we will implement the first-crossing condition simply as
\begin{equation}
\label{eq:crossing}
B < \delta < B + \mu\, \Delta R \;.
\end{equation}
Consequently, the variable $\mu$ will satisfy the constraint $\mu > 0$ rather than 
$\mu > -B'$. 

With the aforementioned modifications, the excursion set peak multiplicity function reads
\begin{align}
\label{fesp}
\fesp(\nu_c) &= \left(\frac{V}{V_*}\right) \frac{1}{\gamma_{\nu\mu}\nu_c}\int_0^\infty 
d\beta\, p(\beta) \\
& \quad\times \int_0^{\infty}d\mu\,\mu\int_0^{\infty} du\, f(u,\alpha=1)\, 
\mathcal{N}(\nu_c+\beta,u,\mu) \nonumber \;,
\end{align}
where $V$ is the Lagrangian volume associated with the TopHat smoothing filter, $V_*  $ 
is the characteristic volume of peaks,  $p(\beta)$ 
is a log-normal distribution, for which we take $\langle \beta \rangle =0.5$ and 
Var$(\beta)=0.25$ as in PSD, and $f(u,\alpha)$ is the slightly modified 
form \citep[see][]{desjacques/crocce/etal:2010} of the original curvature function of 
\cite{bardeen/bond/etal:1986} (see Appendix \S\ref{app:cfunc}). We can now apply Bayes' 
theorem and write $\mathcal{N}(\nu,u,\mu)=\mathcal{N}(\nu,u)\mathcal{N}(\mu|\nu,u)$. The 
integral over $\mu$,
\begin{equation}
\int_0^\infty\!\!d\mu\,\mu\, {\cal N}(\mu|\nu,u) \;,
\end{equation}
is the same as in \cite{musso/sheth:2012} and, therefore, is equal to
\begin{equation}
\bar{\mu} \left[\frac{1+{\rm erf}(\bar{\mu}/\sqrt{2}\Sigma)}{2}
+\frac{\Sigma}{\sqrt{2\pi}\bar{\mu}}e^{-\bar{\mu}^2/2\Sigma^2}\right] \;,
\end{equation}
where
\begin{align}
\bar{\mu} &= 
u\left(\frac{\gamma_{u\mu}-\gamma_1\gamma_{\nu\mu}}{1-\gamma_1^2}\right) 
+(\nu+\beta)\left(\frac{\gamma_{\nu\mu}-\gamma_1\gamma_{u\mu}}{1-\gamma_1^2}\right) \\
\Sigma^2 &= 
\Delta_0^2-\frac{\gamma_{\nu\mu}^2-2\gamma_1\gamma_{\nu\mu}\gamma_{u\mu}
+\gamma_{u\mu}^2}{1-\gamma_1^2} \;.
\end{align}
Substituting this expression into Eq. (\ref{fesp}) and performing numerically the integrals 
over $u$ and $\beta$, we obtain an analytic prediction for the halo mass function without 
any free parameter. Our ESP mass function differs at most by 2 - 3\% over the mass range 
$10^{11} - 10^{15}\hmsun$ from that obtained with the prescription of PSD. Likewise, the 
linear and quadratic local bias parameters are barely affected by our modifications.

\begin{figure*}
\centering 
\resizebox{0.45\textwidth}{!}{\includegraphics{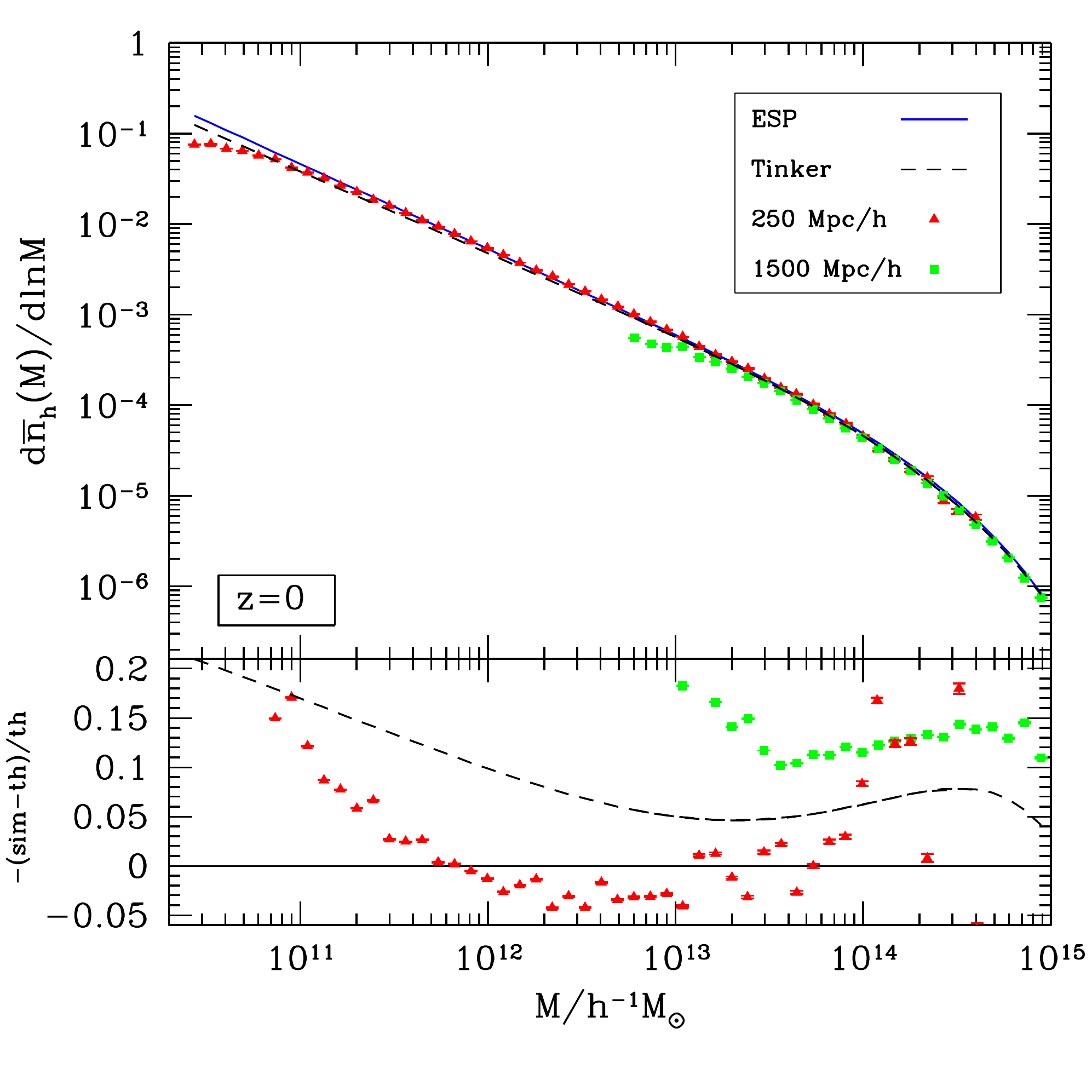}}
\resizebox{0.45\textwidth}{!}{\includegraphics{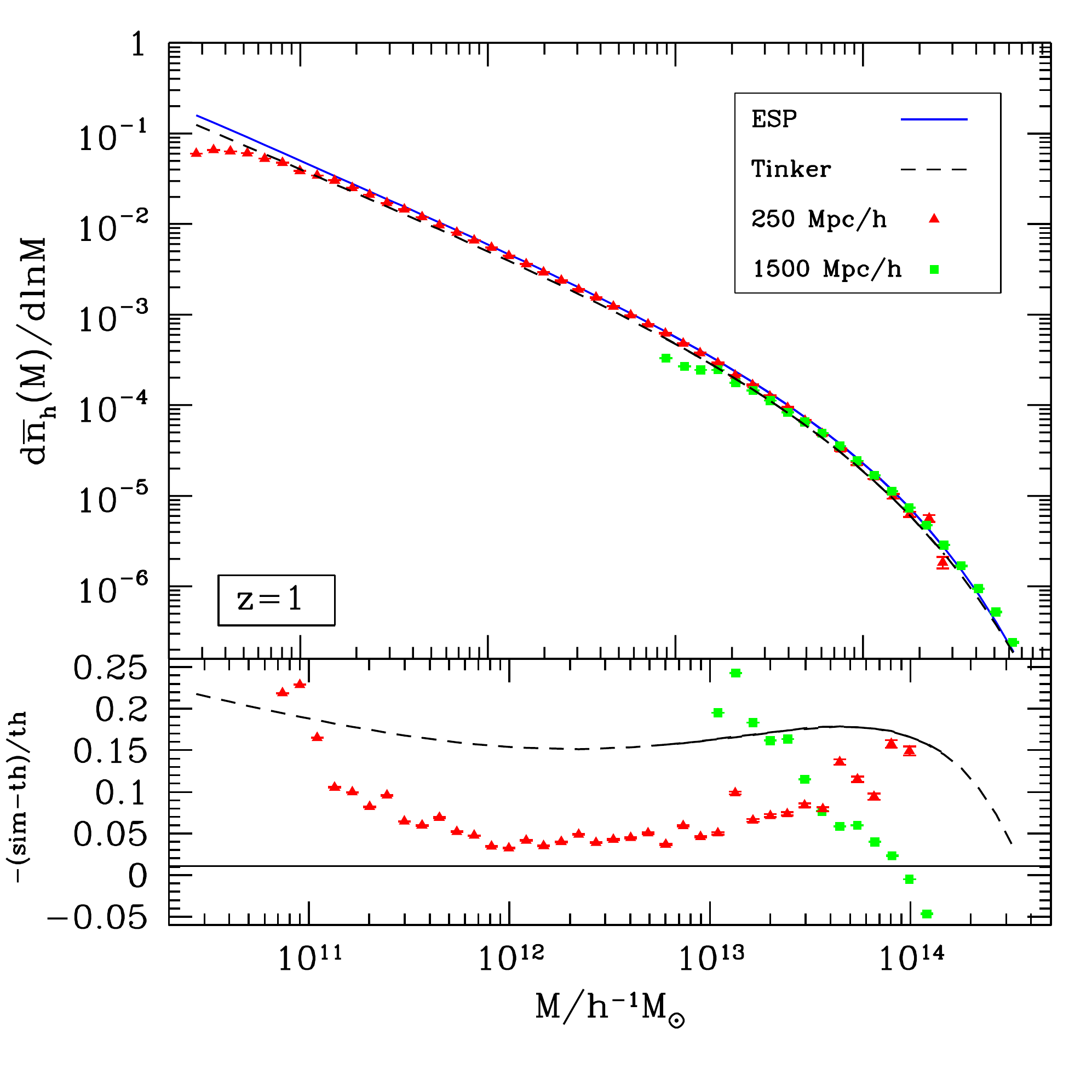}}
\caption{Halo mass function measured from N-body simulation at redshift $z=0$ (left panels) and 
$z=1$ (right panels) with different box sizes as indicated in the figures.  The error bars are 
Poisson. The data is compared to the theoretical prediction Eq.(\ref{eq:nhalo}) based on the 
ESP formalism and the fitting formula of \cite{tinker/kravtsov/etal:2008}. We also show the 
fractional deviation of the \cite{tinker/kravtsov/etal:2008} and the measured halo mass function 
relative to our theoretical prediction.  }
\label{fig:mf}
\end{figure*}

\subsection{Comparison with numerical simulations}

To test the validity of our approach, we compare the ESP mass function with that of haloes
extracted from N-body simulations. For this purpose, we ran a series of N-body
simulations evolving 1024$^3$ particles in periodic cubic boxes of size 1500 and $250\hmpc$.
The particle mass thus is $2.37\times 10^{11}$ and $1.10\times 10^9\hmsun$, respectively. 
The transfer function was computed with {\small CAMB} \citep{lewis/challinor/lasenby:2000} 
assuming parameter values consistent with those inferred by {\small WMAP7} 
\citep{komatsu/smith/etal:2011}: a flat $\Lambda$CDM cosmology with 
$h=0.704$, $\Omega_{\rm m}=0.272$, $\Omega_{\rm b}=0.0455$, $n_s=0.967$ and a normalisation 
amplitude $\sigma_8=0.81$. 
Initial conditions were laid down at redshift $z=99$ with an initial particle displacement
computed at 2nd order in Lagrangian perturbation theory with {\small 2LPTic} 
\citep{crocce/pueblas/scoccimarro:2006}.
The simulations were run using the N-body code {\small GADGET-2} \citep{springel:2005} while 
the halos were identified with the spherical overdensity (SO) halo finder {\small AHF} 
\citep{knollmann/knebe:2009} assuming an overdensity threshold $\Delta_c=200$ constant 
throughout redshift.

In Fig. (\ref{fig:mf}), we compare the simulated halo mass function to the ESP prediction at 
redshift $z=0$ and 1. The latter can be straightforwardly obtained from the multiplicity 
function $\fesp(\nu_c)$ as
\begin{align}
\label{eq:nhalo}
\frac{d \bar{n}_{\rm h}}{d\ln M} &= \frac{\bar{\rho}}{M} \nu_c \fesp(\nu_c,R_s)
\frac{d\log\nu_c}{d\log M} \\
&= -3  R_T\left(\frac{\gamma_{\nu\mu}\nu_c}{\sigma_{0T}}\right) V^{-1}\fesp(\nu_c)
\nonumber \;,
\end{align}
where we used the fact that $\gamma_{\nu\mu}= \sigma_{0T}'$ to obtain the second equality.
The ESP prediction agrees with the simulations at the 10\% level or better from $10^{14}\hmsun$
down to a halo mass $10^{11}\hmsun$, where the correspondence between virialized halos and 
initial density peaks should be rather vague. The abundance of very rare clusters with 
$M>10^{14}\hmsun$ is difficult to predict because of exponential sensitivity to $\delta_c$.
In this respect, it might be more appropriate to work with a critical linear density 
$\delta_c\approx 1.60$ if haloes are defined with a fixed nonlinear threshold $\Delta_c=200$
relative to the mean density \citep[see, e.g.][for a discussion]{barkana:2004,valageas:2009}.

\subsection{Bias parameters}

The bias factors of ESP peaks can be computed using the same formulae as in \cite{desjacques:2013}.
With the additional variable $\mu$, the ``localized'' number density \citep[in the terminology of]
[]{matsubara:2012} can be written as \citep{desjacques/gong/riotto:2013}
\begin{equation}
\nesp({\bf w})=-\left(\frac{\mu}{\gamma_{\nu\mu}\nu_c}\right)\theta_H(\mu)\,n_{\rm pk}(\vy) \;,
\end{equation}
where $\npk$ is the localized number density of BBKS peaks, and 
${\bf w}=(\nu,\eta_i,\zeta_{ij},\mu)\equiv ({\bf y}, \mu)$ is a 11-dimensional vector containing 
all the independent variables of the problem. Therefore, 
\begin{align}
\label{eq:biases}
\sigma_{0T}^i\sigma_{2G}^j b_{ijk}&=
\frac{1}{\bnesp}\int\!\! d^{11}\vw\, \nesp(\vw) H_{ijk}(\nu,u,\mu) P_1(\vw) 
\nonumber \\
\sigma_{1G}^{2k}\chi_{k0}&=
\frac{(-1)^k}{\bnesp}\int\!\! d^{11}\vw\, \nesp(\vw) 
L^{(1/2)}_k\!\!\left(\frac{3\eta^2}{2}\right)P_1(\vw) \nonumber \\
\sigma_{2G}^{2k}\chi_{0k}&=
\frac{(-1)^k}{\bnesp}\int\!\! d^{11}{\bf w}\, \nesp(\vw)
L^{(3/2)}_k\!\!\left(\frac{5\zeta^2}{2}\right)P_1(\vw) \;.
\end{align}
Here, $P_1(\vw)$ is the 1-point probability density
\begin{align}
P_1(\vw)d^{11}\vw &= \mathcal{N}(\nu,u,\mu) d\nu dud\mu \times \chi^2_3(3\eta^2)d(3\eta^2) \\
& \qquad \times\chi^2_5(5\zeta^2)d(5\zeta^2)\times P({\rm angles}) \nonumber \;,
\end{align}
where $H_{ijk}(\nu,u,\mu)$ are trivariate Hermite polynomials and $\chi_k^2(x)$ is a 
$\chi^2$-distribution with $k$ degrees of freedom (d.o.f.). The probability density $P$(angles)
\citep[which was missing~\footnote{We thank Marcello Musso for pointing this out to us.} in]
[]{desjacques:2013} represents the probability distribution of the 5 remaining degrees of freedom.
Since they are all angular variables, they do not generate bias factors because the peak
(and halo) overabundance can only depend on scalar quantities 
\citep[e.g.][]{catelan/matarrese/porciani:1998,mcdonald/roy:2009}.

The behaviour of the bias factors $b_{ij0}$ and $\chi_{kl}$ as a function of halo mass is similar
to that seen in Fig.1 of \cite{desjacques:2013}. The bias factors $b_{ijk}$ with $k\geq 1$ weight
the contributions of $\mu^k$ terms to the clustering of ESP peaks that are proportional to 
derivatives of the tophat filter w.r.t. the filtering scale $R_T$. Similar contributions appear 
in the clustering of thresholded regions \citep{matsubara:2012,ferraro/smith/etal:2012} since 
their definition also involve a first-crossing condition.

The effective bias expansion takes the form \citep{desjacques:2013,desjacques/gong/riotto:2013}
\begin{align}
\label{eq:dpk}
\dpk(\vx) &= 
\sigma_{0T} b_{100}\nu(\vx)+\sigma_{2G} b_{010} u(\vx) + b_{001}\mu(\vx) \\
&\quad + \frac{1}{2}\sigma_{0T}^2 b_{200}\nu^2(\vx) +\sigma_{0T}\sigma_{2G} b_{110}\nu(\vx) u(\vx) 
\nonumber \\
&\quad +\frac{1}{2}\sigma_{2G}^2 b_{020} u^2(\vx) + \frac{1}{2}b_{002}\mu^2(\vx) \nonumber \\ 
&\quad +\sigma_{1G}^2\chi_{10}\eta^2(\vx) +\sigma_{2G}^2\chi_{01}\zeta^2(\vx) \nonumber \\
&\quad + \sigma_{0T} b_{101}\nu(\vx)\mu(\vx) +\sigma_{2G} b_{011}u(\vx)\mu(\vx) + \cdots 
\nonumber
\end{align}
Here, the rule of thumb is that one should ignore all the terms involving zero-lag moments in the
computation of $\la\dpk(\vx_1)\dots\dpk(\vx_N)\ra$ in order to correctly predict the $N$-point 
correlation function, as demonstrated explicitly in \cite{desjacques:2013}.
The appearance of rotationally invariant quantities is, again, only dictated by the scalar nature 
of the peak overabundance. The variables of interest here are
\begin{align}
\label{eq:chivars}
\eta^2(\vx) &= \frac{1}{\sigma_{1G}^2}\left(\nabla\delta\right)^2\!(\vx) \\
\zeta^2(\vx) &= \frac{3}{2\sigma_{2G}^2}
{\rm tr}\!\biggl[\Bigl(\partial_i\partial_j\delta-\frac{1}{3}\delta_{ij}\nabla^2\delta\Bigr)^2\biggr]\!(\vx)
\nonumber \;,
\end{align}
so that $3\eta^2(\vx)$ and $5\zeta^2(\vx)$ are $\chi^2$-distributed with 3 and 5 d.o.f., respectively.

\section{Biases from cross-correlation: extension to $\chi^2$ variables}
\label{sec:CHICROSS}

In this Section, we will demonstrate that the bias factors $\chi_{ij}$ can be measured
with a one-point statistics.
We will test our method on density peaks of a Gaussian random field before applying it
to dark matter halos.

\subsection{Bias factors $b_{ijk}$: Hermite polynomials}

\cite{musso/paranjape/sheth:2012} showed that the bias factors of discrete tracers 
(relative to the mass density $\delta$) can be computed from one-point measurements 
rather than computationally more expensive $n$-point correlations. Their idea was 
implemented by \cite{paranjape/sheth/desjacques:2013,paranjape/sefusatti/etal:2013} 
to haloes extracted from N-body simulations in order to test the predictions of the
ESP formalism. Namely, haloes were traced back to their ``proto-halo'' patch (since
one is interested in measuring Lagrangian biases) in the initial conditions, the 
linear density field was smoothed on some ``large scale'' $R_l$ and the quantity 
$H_n(\nu_l=\delta_l/\sigma_{0l})$ was computed (for $n=1,2$ only) at the location 
of each proto-halo. The average of $H_n(\nu_l)$ over all proto-haloes reads
\begin{equation}
\label{eq:ensHn}
\frac{1}{N}\sum_{i=1}^N H_n(\nu_l) =
\int_{-\infty}^{+\infty}\!\!d\nu_l\,{\cal N}(\nu_l) 
\bigl\langle 1+\delta_h \bigl\lvert \nu_l\bigr\rangle H_n(\nu_l) \;,
\end{equation}
where $\delta_h$ is the overdensity of proto-haloes. This expression assumes that 
the first-crossing condition can be implemented through a constraint of the form 
Eq.(\ref{eq:crossing}), so that $P(\nu_l)$ is well approximated by a Gaussian 
\citep{musso/paranjape/sheth:2012}.
For the ESP peaks considered here, this ensemble average reads
\begin{align}
\frac{1}{\bar{n}_{\rm ESP}}\int_{-\infty}^{+\infty}\!\!d\nu_l&\,{\cal N}(\nu_l) 
\bigl\langle n_{\rm ESP}\bigl\lvert \nu_l\bigr\rangle H_n(\nu_l) \\
&= \frac{1}{\bnesp}\int\!\!d^{11}\vw\,\nesp(\vw)\,\left(-\epsilon_\nu\right)^n 
\nonumber \\ &\qquad \times
\left(\frac{\partial}{\partial\nu}+\frac{\epsilon_u}{\epsilon_\nu}\frac{\partial}{\partial u}
+\frac{\epsilon_\mu}{\epsilon_\nu}\frac{\partial}{\partial\mu}\right)^n P_1(\vw)
\nonumber \;.
\end{align}
Here, $\epsilon_X$ denotes the cross-correlation between $\nu_l$ and the variables
$X=(\nu,u,\mu)$ defined at the halo smoothing scale. The right-hand side reduces
to a sum of $n$th-order bias factors $b_{ijk}$ weighted by products of $\epsilon_\nu$,
$\epsilon_u$ and $\epsilon_\mu$. Relations between bias factors of a given order 
\citep[which arise owing to their close connection with Hermite polynomials, see e.g.]
[]{musso/paranjape/sheth:2012} can then be used to extract a measurement of each
$b_{ijk}$. 

\begin{figure*}
\centering
\includegraphics[width=\linewidth]{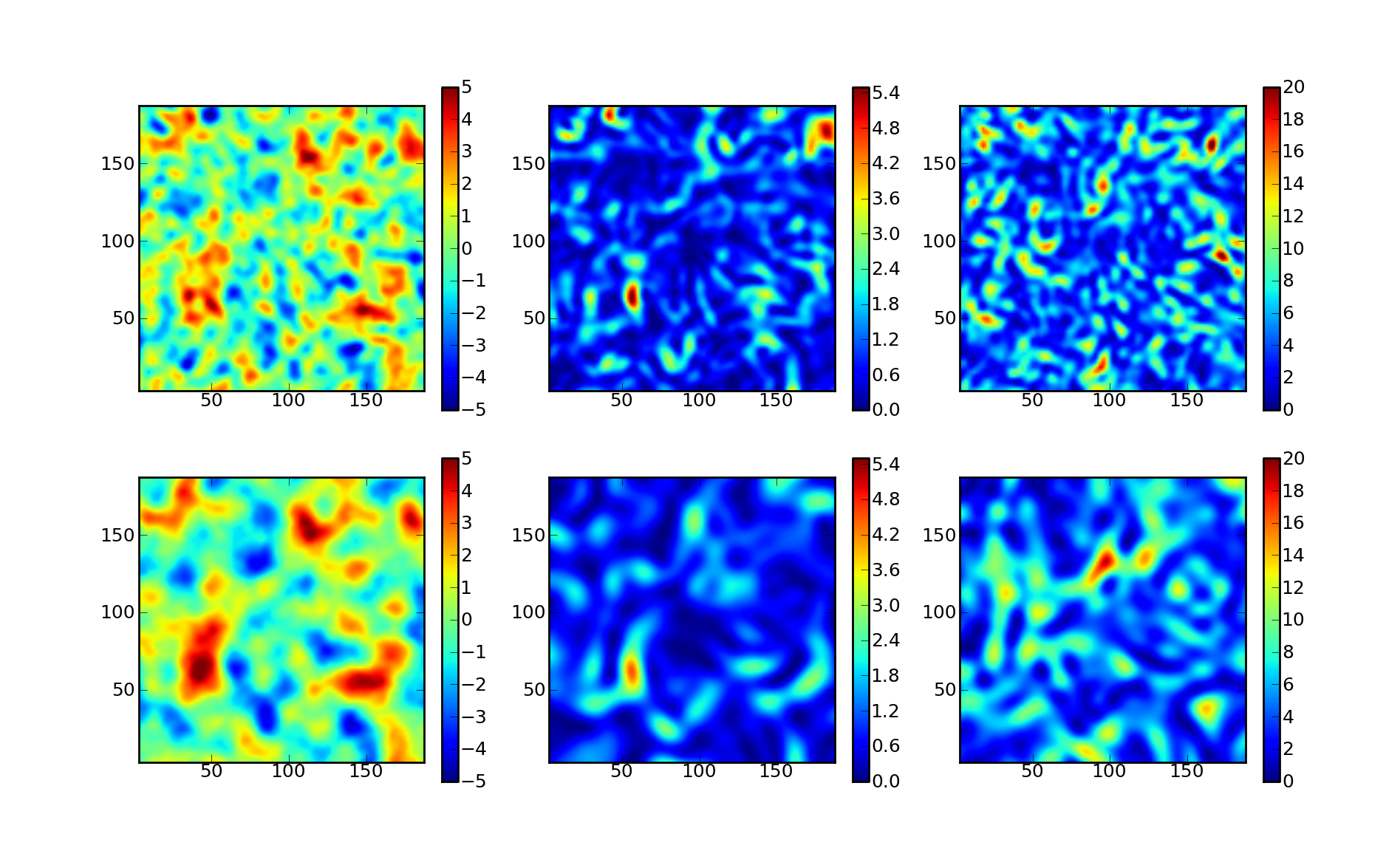}
\caption{Sections for $\nu_l$,  $ 3  \eta_l^2  $ and $ 5 \zeta_l^2 $ (from left to 
right). A filtering scale of $R_l=5$ and 10$\hmpc$ is used for the first and second 
row, respectively. Note that a tophat kernel is applied for $\nu_l$, while a Gaussian 
window is used for  $\eta_l^2$ and $\zeta_l^2$.  In each panel, the dimension of 
the section is 200$\times$200 ${\it h}^{-2}$Mpc$^2$.}
\label{fig:section_biastype123_RG5_10}
\end{figure*}

Before we generalize this approach to the chi-squared bias factors $\chi_{ij}$, we
emphasize that, in this cross-correlation approach, the smoothing scale $R_l$ can 
take any value as long as it is distinct from the halo smoothing scale. 
\cite{paranjape/sefusatti/etal:2013} chose $R_l\gg R_s$ in the spirit of the 
peak-background split but this requirement is, in fact, not necessary as long as the
correlation between the two scales is taken into account. In any case, we will stick 
with the notation $R_l$ for convenience.

\subsection{Bias factors $\chi_{ij}$:  Laguerre polynomials}

The approach presented above can be generalised to $\chi^2$ distributions. 
The main difference is the appearance of Laguerre polynomials $L_n^{(\alpha)}$. 
Consider for instance the $\chi^2$-quantity $3\eta^2$ smoothed at the scale $R_l$, 
i.e. $3\eta_l^2$.
In analogy with Eq.(\ref{eq:ensHn}), the ensemble average of $L_n^{(1/2)}(3\eta_l^2)$ 
at the peak positions is
\begin{align}
\label{eq:ln3}
\frac{1}{N}\sum_{i=1}^N 
L_n^{(1/2)}\!\left(\frac{3\eta_l^2}{2}\right) &=
\int_0^\infty\!\!d(3\eta_l^2)\, \chi_3^2(3\eta_l^2) \\
& \qquad \times \bigl\langle 1+\delta_h \bigl\lvert 3\eta_l^2\bigr\rangle
L_n^{(1/2)}\!\!\left(\frac{3\eta_l^2}{2}\right) \nonumber \;.
\end{align}
The conditional average $\bigl\langle 1 + \delta_h\bigl\lvert 3\eta_l^2\bigr\rangle$
reads
\begin{align}
\bigl\langle 1 + \delta_h\bigl\lvert 3\eta_l^2\bigr\rangle &=
\frac{1}{\bnesp}\int\!\! d^{11}\vw\, \nesp(\vw) P_1\bigl(\vw\bigl\lvert3\eta_l^2\bigr) \\
&=\frac{1}{\bnesp}\int\!\!du d\nu d\mu\, {\cal N}(\nu,u,\mu) \nonumber \\ 
&\qquad \times\int\!\!d(3\eta^2)\,\chi_3^2(3\eta^2|3\eta_l^2) 
\int\!\!d(5\zeta^2)\,\chi_5^2(5\zeta^2) \nonumber \\
&\qquad \times\int\!\!d(\text{angles})\,P(\text{angles})\,\nesp(\vw) \nonumber \;.
\end{align}
We substitute this relation into Eq.(\ref{eq:ln3}) and begin with the integration over
the variable $3\eta_l^2$.

We use the following relation \citep[which can be inferred from 
Eq.(7.414) of][]{gradshteyn/ryzhik:1994}
\begin{equation}
\int_0^\infty\!\!dx\,e^{-x} x^{j+\alpha}L_n^{(\alpha)}\!(x)
=\frac{(-1)^n}{n!}\frac{j!\,\Gamma(j+\alpha+1)}{(j-n)!} \;.
\end{equation}
With the aid of this result and on expanding the conditional $\chi^2$-distribution 
$\chi_3^2(3\eta^2|3\eta_l^2)$ in Laguerre polynomials (see Appendix \S\ref{app:chidist} 
for details), we obtain
\begin{align}
\label{eq:seriesL}
\int_0^\infty\!\!& d(3\eta_l^2)\,\chi_3^2(3\eta_l^2)\, 
L_n^{(1/2)}\!\left(\frac{3\eta_l^2}{2}\right) \,
\chi_3^2(3\eta^2|3\eta_l^2) \\
&=\frac{(-1)^n}{n!}\frac{1}{\Gamma(3/2)}
\left(\frac{3\eta^2}{2}\right)^\alpha
\frac{e^{-3\eta^2/2(1-\epsilon^2)}}{2\left(1-\epsilon^2\right)^{\alpha+1}} 
\nonumber \\ &\qquad \times 
\sum_{j=0}^\infty\frac{j!}{(j-n)!}\left(\frac{-\epsilon^2}{1-\epsilon^2}\right)^j
L_j^{(1/2)}\!\left[\frac{3\eta^2}{2(1-\epsilon^2)}\right] \nonumber \;.
\end{align}
For simplicity, let us consider the case $n=0,1$ solely. 
For $n=0$, the sum simplifies to
\begin{multline}
\sum_{j=0}^\infty \left(\frac{-\epsilon^2}{1-\epsilon^2}\right)^j
L_j^{(1/2)}\!\left[\frac{3\eta^2}{2(1-\epsilon^2)}\right] \\
= \left(1-\epsilon^2\right)^{3/2}
\exp\!\left[\left(\frac{\epsilon^2}{1-\epsilon^2}\right)\frac{3\eta^2}{2}\right]\;,
\label{eq:n=0}
\end{multline}
and the integral Eq.(\ref{eq:seriesL}) ($L_0^{(1/2)}(3\eta_l^2/2)\equiv 1$) is 
trivially equal to $\chi_3^2(3\eta^2)$ (as it should be, since we are essentially 
marginalizing over $3\eta_l^2$). 

For $n\geq 1$, the sum can be evaluated upon taking suitable derivatives of the 
right-hand side of Eq.(\ref{eq:n=0}), which indeed is a generating function for
the Laguerre polynomials $L_n^{(1/2)}$. For $n=1$, a little algebra leads to
\begin{multline}
\sum_{j=0}^\infty j \left(\frac{-\epsilon^2}{1-\epsilon^2}\right)^{j-1}
L_j^{(1/2)}\!\left[\frac{3\eta^2}{2(1-\epsilon^2)}\right] \\
= \left(1-\epsilon^2\right)^{5/2} L_1^{(1/2)}\!\left(\frac{3\eta^2}{2}\right)
\exp\left[\left(\frac{\epsilon^2}{1-\epsilon^2}\right)\frac{3\eta^2}{2}\right]\;.
\end{multline}
Hence, Eq.~(\ref{eq:seriesL}) with $n=1$ equals 
$\epsilon^2 L_1^{(1/2)}(3\eta^2/2)\chi_3^2(3\eta^2)$. Performing the remaining 
integrals over $\nu$, $u$, $\mu$ and $5\zeta^2$ (the integral over the angles is
trivially unity) and taking into account the ESP peak constraint through the 
multiplicative factor $\nesp(\vw)$, Eq.(\ref{eq:ln3}) simplifies to
\begin{equation}
\int_0^\infty\!\!d(3\eta_l^2)\, 
\chi_3^2(3\eta_l^2)\,
\bigl\langle 1+\delta_h \bigl\lvert 3\eta_l^2\bigr\rangle\,
L_1^{(1/2)}\!\!\left(\frac{3\eta_l^2}{2}\right)
= - \epsilon^2 \sigma_1^2 \chi_{10}
\end{equation}
For the variable $3\eta^2$, the cross-correlation coefficient $\epsilon$ is
\begin{equation}
\epsilon^2 \equiv \frac{\bigl\langle\eta^2\eta_l^2\bigr\rangle-
\bigl\langle\eta^2\bigr\rangle\bigl\langle\eta_l^2\bigr\rangle}
{\sqrt{
\left(\bigl\langle\eta^4\bigr\rangle-\bigl\langle\eta^2\bigr\rangle^2\right)
\left(\bigl\langle\eta_l^4\bigr\rangle-\bigl\langle\eta_l^2\bigr\rangle^2\right)
}}
=\left(\frac{\sigma_{1\times}^2}{\sigma_{1s}\sigma_{1l}}\right)^2 \;,
\end{equation} 
which we shall denote as $\epsilon_1$ in what follows. Furthermore,
\begin{align}
\sigma_{n\times}^2 = 
\frac{1}{2\pi^2} \int_0^\infty\!\!dk\,k^{2(n+1)}\,P(k) W_G(k R_s) W_G(k R_l)
\end{align}
designates the splitting of filtering scales, i.e. one filter is on scale 
$R_s$ while the second is on scale $R_l$. It should be noted that, unlike 
$\sigma_{1{\rm\small X}}$ defined in Eq.(\ref{eq:sigma1x}), both filtering 
kernels are Gaussian.

The derivation of the bias factors $\chi_{0k}$ associated with the quadratic variable
$\zeta^2$ proceeds analogously. In particular,
\begin{equation}
\int_0^\infty\!\!d(5\zeta_l^2)\, 
\chi_5^2(5\zeta_l^2)\,
\bigl\langle 1+\delta_h \bigl\lvert 5\zeta_l^2\bigr\rangle\,
L_1^{(3/2)}\!\!\left(\frac{5\zeta_l^2}{2}\right)
= - \epsilon^2 \sigma_2^2 \chi_{01} \;.
\end{equation}
Here, the cross-correlation coefficient is 
$\epsilon=\sigma_{2\times}^2/(\sigma_{2s}\sigma_{2l})\equiv \epsilon_2$. Note that, 
in both cases, the cross-correlation coefficient drops very rapidly as $R_l$ moves 
away from $R_s$ for realistic CDM power spectra. In addition, one could in principle 
choose $R_l< R_s$ (if there is enough numerical resolution) to measure $\chi_{ij}$.

\section{Test with numerical simulations}
\label{sec:SIMS}

In this Section, we first validate our predictions based on peaks of Gaussian 
random fields with measurements extracted from random realizations of the Gaussian 
linear density field, and then move on to calculate $\chi_{10}$ and $\chi_{01}$
for $M\bsim M_\star$ haloes, where $ M_\star$ is the characteristic mass of the haloes. 

\subsection{Peaks of Gaussian random fields}

\begin{figure*}
\centering
\resizebox{0.45\textwidth}{!}{\includegraphics{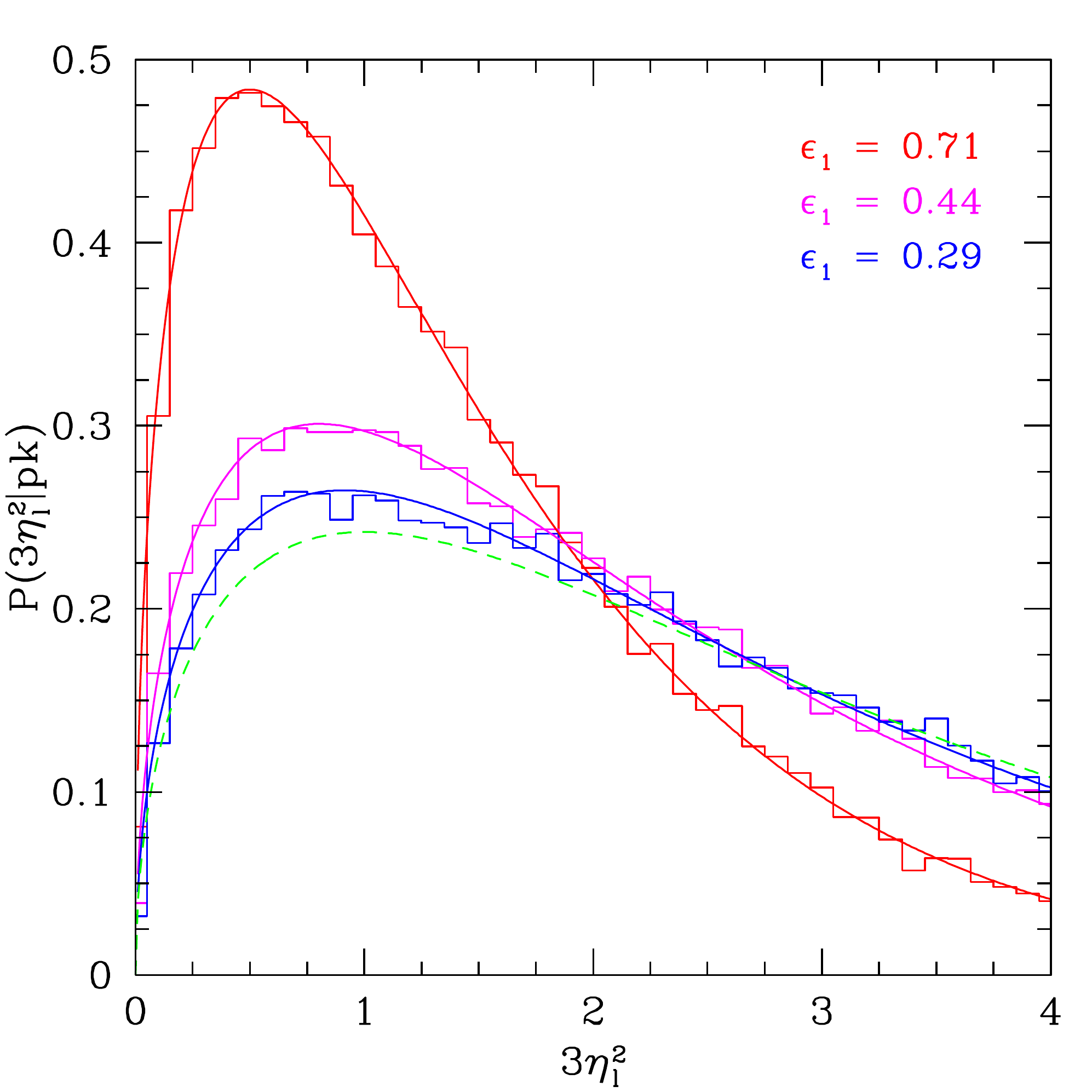}}
\resizebox{0.45\textwidth}{!}{\includegraphics{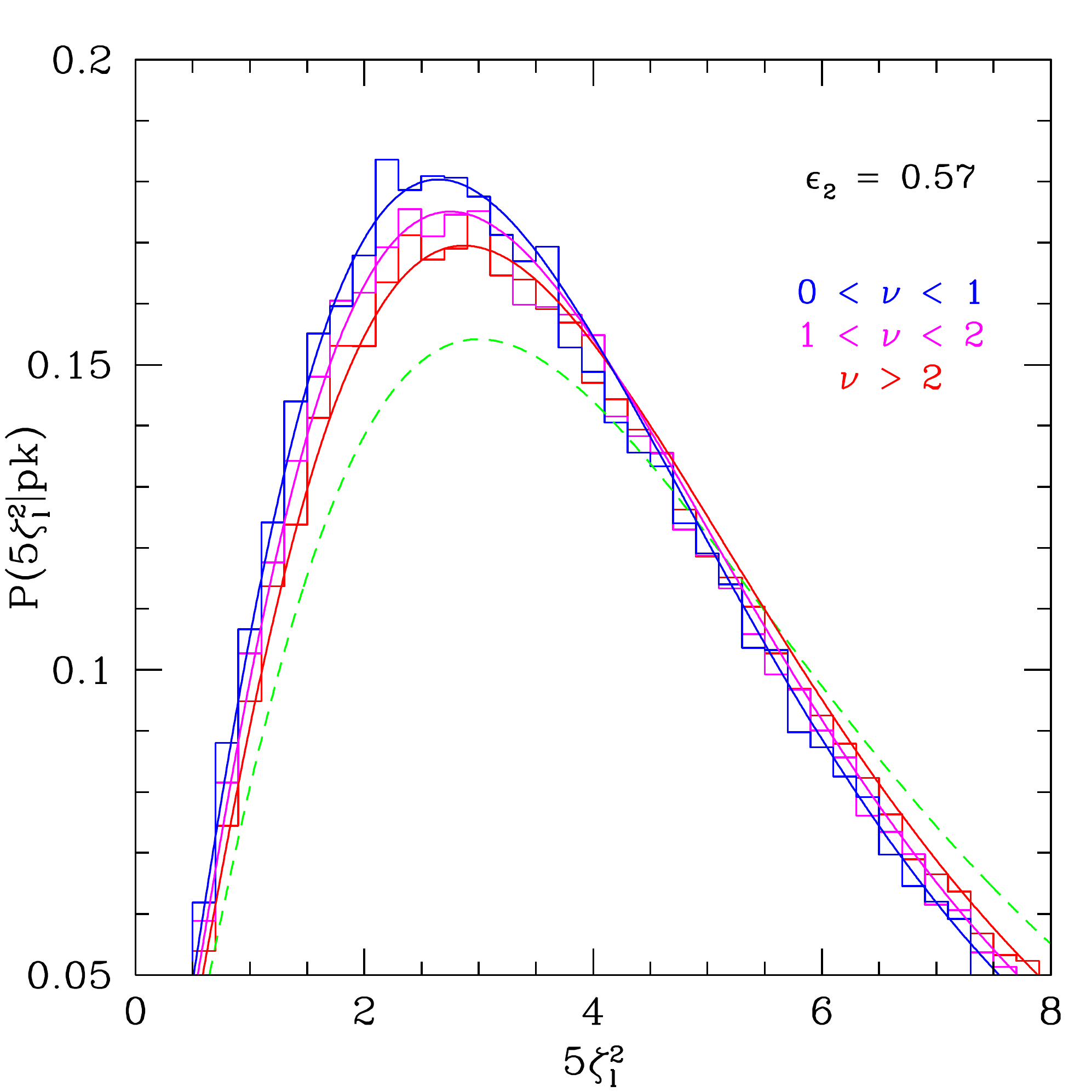}}
\caption{Conditional probability distribution for the variables $3\eta_l^2$ (left panel) 
and $5\zeta_l^2$ (right panel) measured at the position of maxima of the linear density 
field smoothed with a Gaussian filter on scale $R=5\hmpc$. {\it Left panel}: histrograms 
indicate the results for $R_l=10$, 15 and $20\hmpc$, which leads to $\epsilon_1=0.71$, 
0.44 and 0.29 as quoted on the figure.
{\it Right panel}: histograms show the results for a fixed $R_l=10\hmpc$ (which implies
$\epsilon_2=0.57$) but several peak height intervals. In all cases, the solid curves are 
the theoretical prediction (see text) whereas the dashed (green) curves represents the 
unconditional distribution $\chi_k^2(y)$.}
\label{fig:chipk}
\end{figure*}

We generate random realizations of the Gaussian, linear density field with a power 
spectrum equal to that used to seed the N-body simulations described above. 
To take advantage of FFTs, we simulate the linear density field in periodic, cubic 
boxes of side 1000$\hmpc$. The size of the mesh along each dimension is 1536. 
We smooth the density field on scale $R_s=5\hmpc$ with a tophat filter and find 
the local maxima by comparing the density at each grid point with its 26 neighbouring 
values. 

We then smooth the density field on the larger scales $R_l=10$, 15 and 20$\hmpc$ with
a Gaussian filter and compute
\begin{align}
\eta_l^2 &= \frac{1}{\sigma_{1l}^2}(\nabla\delta_l)^2 \\
\zeta_l^2 &= \frac{3}{2\sigma_{2l}^2}
{\rm tr}\!\biggl[\Bigl(\partial_i\partial_j\delta_l
-\frac{1}{3}\delta_{ij}\nabla^2\delta_l\Bigr)^2\biggr]\;.
\end{align}
These density fields with derivatives sensitively depend on the smoothing scales 
used. To illustrate this we show in Fig.~\ref{fig:section_biastype123_RG5_10} 
sections of $ \nu_l$,  $3 \eta_l^2  $ and $ 5  \zeta_l^2 $. 
The sections, each of which of dimensions 200$\times$200 ${\rm {\it h}^{-2}Mpc^2}$, 
were generated at $z=99$ with the same random seed.
The first row corresponds to $R_s=5 \hmpc$, whereas the second row displays results 
on the filtering scale $R_l= 10\hmpc$.
We note that, for the normalized density field $\nu_l$, an increase in the smoothing 
scale washes out the small scale features, but the large scale pattern remains. 
For the quadratic variable $\eta_l^2$ however, the resemblance between the features 
at the small and large filtering scale is tenuous. This is even worse for $\zeta_l^2$.

Compared to $\nu_l$, the fields $\eta_l^2$ and $\zeta_l^2$ have one and two additional
derivatives which give rise to an effective window function whose isotropic part is 
given by
\begin{equation}
W_{\rm eff}(k,R) = k^n e^{ -( k R)^2 /2 }, 
\end{equation}
where $n=0$, 1 and 2 are for $\nu_l$, $\eta_l^2$ and $\zeta_l^2$, respectively. 
For $n=0$, the window becomes narrower as $R_l$ increases, yet remains unity for 
wavenumbers $k\lesssim 1/R_l$.  $W_{\rm eff} $ reaches a maximum at $\sqrt{n} /R $. 
Hence, for $n=1$ and 2, $W_{\rm eff }$ selects predominantly wavemodes with $k\sim 1/R $. 
Consequently, since in a Gaussian random field the wavemodes at different scales are 
uncorrelated, patterns in the fields $\eta_l^2$ and $\zeta_l^2$ can change drastically 
as $R_l$ varies. 
This effect expected to be most significant for $n=2$, i.e. $\zeta_l^2$.

For each local density maxima, we store the peak height $\nu$ as well as the value of 
$\eta_l^2$ and $\zeta_l^2$ at the peak position. 
The left panel of Fig.\ref{fig:chipk} displays as histograms the resulting probability 
distribution $P(3\eta_l^2|{\rm pk})$ for three different values of $R_l=10$, 15 and 
20$\hmpc$. 
The solid curves represent the theoretical prediction Eq.(\ref{eq:chixy1}) with 
$x=\la 3\eta^2|{\rm pk}\ra = 0$ and 
$\epsilon_1=0.71$, 0.44 and 0.29 (from the smallest to largest $R_l$) as was measured from 
the random realizations. The dashed curve is the unconditional $\chi^2$-distribution with
3 degrees of freedom. The theory gives excellent agreement with the simulations. Note 
also that we did not find any evidence for a dependence on the peak height, as expected
from the absence of a correlation between $\nu$ and $\eta_l^2$. 
The right panel of Fig.\ref{fig:chipk} shows results for $\zeta_l^2$. Here however, since 
the cross-correlation coefficient diminishes very quickly when $R_l$ even slightly departs
from $R_s$, we show result for $R_l=10\hmpc$ only, which corresponds to $\epsilon_2=0.57$.
In addition, because one should expect a dependence of the shape of the density profile
around peaks to the peak height, we consider three different ranges of $\nu$ as indicated
on the figure. The solid curves indicate the theoretical prediction Eq.(\ref{eq:chixy1})
with $\epsilon_2=0.57$ and $x=\la 5\zeta^2|{\rm pk}\ra$, where
\begin{equation}
\la 5\zeta^2|{\rm pk}\ra = 
-2\partial_\alpha\ln\int_{\nu_{\rm min}}^{\nu_{\rm max}}d\nu\,
G_0^{(\alpha)}\!(\gamma_1,\gamma_1\nu)\;.
\end{equation}
Here, $G_0^{(\alpha)}$ is the integral of $f(u,\alpha)$ over all the allowed peak curvatures.
The average $\la 5\zeta^2|{\rm pk}\ra$ increases with the peak height to reach 5 in the 
limit $\nu\to\infty$. The figure shows a clear deviation from the unconditional distribution
$\chi_5^2(5\zeta_l^2)$ (shown as the dashed curve) and a dependence on $\nu$ consistent
with theoretical predictions.

\subsection{Dark matter haloes}

Having successfully tested the theory against numerical simulations of Gaussian peaks, 
we will now attempt to estimate the bias factors $\chi_{10}$ and $\chi_{01}$ associated
with dark matter haloes. For this purpose, we first trace back all dark matter particles 
belonging to virialized haloes at redshift $z=0$ to their initial position at $z=99$. We 
then compute the center-of-mass positions of these Lagrangian regions and assume that they 
define the locations of proto-haloes. We can now proceed as for the Gaussian peaks and 
compute $\nu$, $\eta_l^2$ and $\zeta_l^2$ at the position of proto-haloes. 

The quadratic bias factors $\chi_{10}$ and $\chi_{01}$ could be in principle computed 
analogously to \cite{paranjape/sefusatti/etal:2013}, i.e. by stacking measurements of 
$\eta_l^2$ and $\zeta_l^2$ at the locations of proto-haloes~:
\begin{equation}
\sigma_{1s}^2 \hat{\chi}_{10} = 
-\frac{1}{N\epsilon_1^2}\sum_{i=1}^N 
L_1^{(1/2)}\!\left(\frac{3\eta_l^2}{2}\right)
\end{equation}
and
\begin{equation}
\sigma_{2s}^2 \hat{\chi}_{01} = 
-\frac{1}{N\epsilon_2^2}\sum_{i=1}^N 
L_1^{(3/2)}\!\left(\frac{5\zeta_l^2}{2}\right) \;.
\end{equation}
Here, $N$ is the number of halos, $s$ designates smoothing at the halo mass scale with a 
Gaussian filter $W_G$ on scale $R_G(R_T)$, whereas $l$ designates Gaussian smoothing at 
the large scale $R_l$. However, because the cross-correlation 
coefficient is fairly small unless $R_l$ is very close to $R_G$, we decided to compute 
$\chi_{10}$ and $\chi_{01}$ by fitting the probability distribution 
$P(3\eta_l^2|{\rm halo})$ and $P(5\zeta_l^2|{\rm halo})$ with the conditional 
$\chi^2$-distribution $\chi_k^2(y|x)$. Namely,
\begin{align}
\label{eq:chiestimate}
\sigma_{1s}^2\hat{\chi}_{10} &= \frac{1}{2}\left(\la 3\eta^2|{\rm halo}\ra -3\right) \\
\sigma_{2s}^2\hat{\chi}_{01} &= \frac{1}{2}\left(\la 5\zeta^2|{\rm halo}\ra -5\right)
\nonumber \;,
\end{align}
where $\la 3\eta^2|{\rm halo}\ra$ and $\la 5\zeta^2|{\rm halo}\ra$ are the best-fitting
values obtained for $x$. We used measurements obtained at the smoothing scale 
$R_l=10\hmpc$ only to maximize the signal.

\begin{figure}
\centering
\resizebox{0.45\textwidth}{!}{\includegraphics{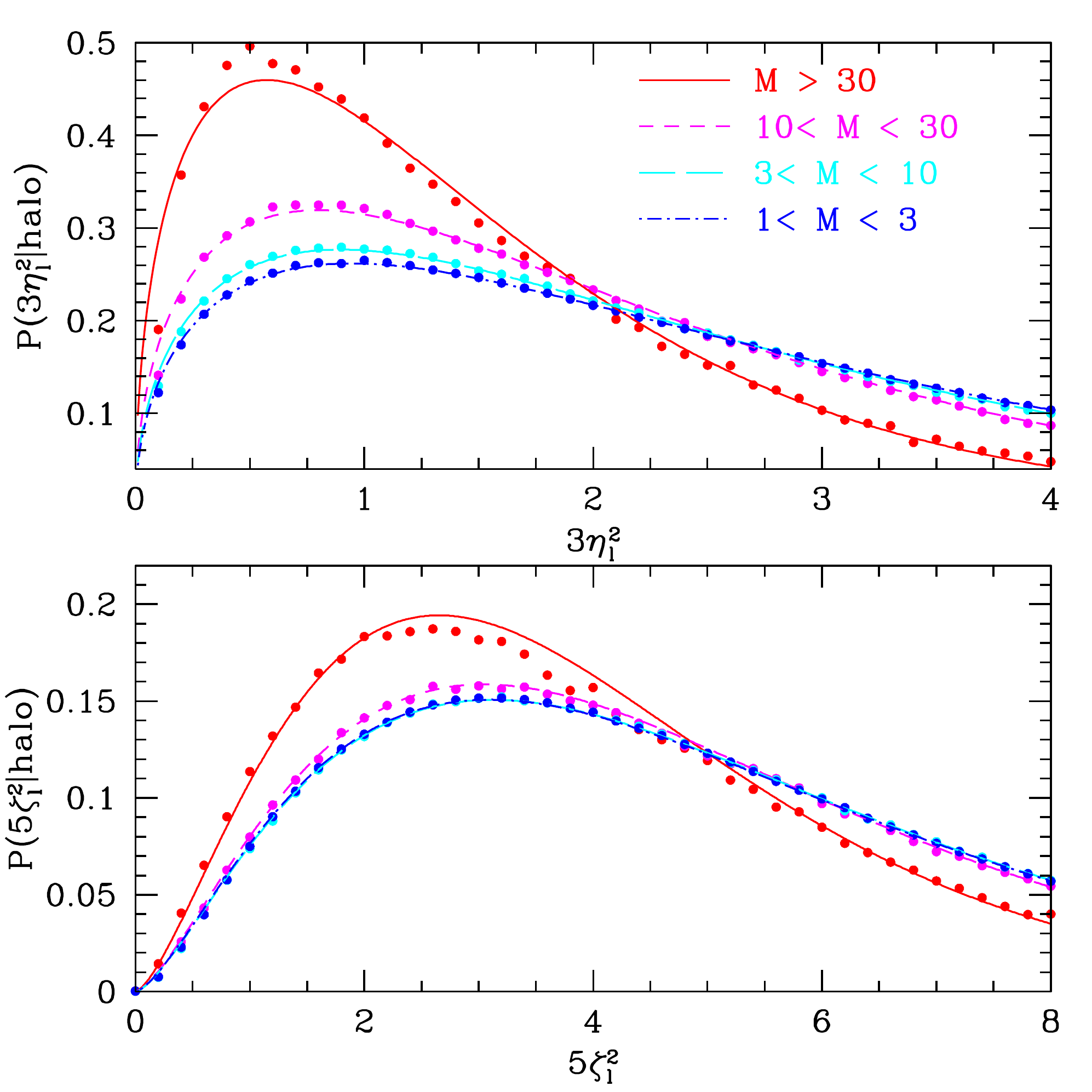}}
\caption{Conditional probability distribution for $3\eta_l^2$ (top panel) and $5\zeta_l^2$ 
(bottom panel) measured at the center-of-mass position of proto-halos. The filter is Gaussian
with $R_l=10\hmpc$. The various curves show the best-fit theoretical predictions 
for the halo mass bins considered here. Halo mass range are in unit of $10^{13}\hmsun$.
Poisson errors are much smaller than the size of the data points and, therefore, do not
show up on the figure.}
\label{fig:chiha}
\end{figure}

To predict the value of $R_G$ given $R_T$, 
we followed PSD and assumed that $R_G(R_T)$ can be 
computed through the requirement that $\la\delta_G|\delta_T\ra = \delta_T$.
This yields a prediction for the value of the cross-correlation coefficients $\epsilon_1$
and $\epsilon_2$ as a function of halo mass, which we can use as an input to $\chi_k^2(y|x)$ 
and only fit for $x$. However, we found that using the predicted $\epsilon_1$ leads to 
unphysical (negative) values for $x$ when one attempts to fit $P(3\eta_l^2|{\rm halo})$. 
Therefore, we decided to proceed as follows: 
\begin{enumerate}
\item Estimate both $\epsilon_1$ and $x=\la 3\eta^2|{\rm halo}\ra$ 
by fitting the model $\chi_3^2(y|x;\epsilon_1)$ to the measured $P(3\eta_l^2|{\rm halo})$.
\item Compute $\epsilon_2$ assuming that the same $R_G$ enters the spectral moments.
\item Estimate $x=\la 5\zeta_l^2|{\rm halo}\ra$ by fitting the theoretical model 
$\chi_5^2(y|x;\epsilon_2)$ to the simulated $P(5\zeta_l^2|{\rm halo})$.
\end{enumerate}
We considered data in the range $0<3\eta_l^2<8$ and $0<5\zeta_l^2<12$ and gave equal 
weight to all the measurements (assuming Poisson errors does not affect our results 
significantly). Table \ref{table:table1} summarizes the best-fitting values 
obtained for four different halo bins spanning the mass range $10^{13} - 10^{15}\hmsun$,
whereas the measured probability distributions together with the best-fit models are 
shown in Fig.\ref{fig:chiha}. The data is reasonably well described by a conditional 
$\chi^2$-distribution, but the fit is somewhat poorer when the cross-correlation 
coefficient is close to unity. 

\begin{table}
\caption{Best-fit parameter values as a function of halo mass. The 
latter is in unit of $10^{13}\hmsun$. Note that we also list the values of
$\epsilon_2$ even though it is not directly fitted to the data (see text
for details).}
\begin{center}
\begin{tabular}{ccccc} \hline\hline
Halo mass &  $\la 3\eta^2|{\rm halo}\ra$ & $\epsilon_1$ & $\la 5\zeta^2|{\rm halo}\ra$ 
& ($\epsilon_2$) \\ 
\hline 
$M > 30$ & 0.71 & 0.80 & 2.98 & (0.70) \\
$10 < M < 30$ & 1.24 & 0.66 & 4.49 & (0.52) \\
$3 < M < 10$ & 1.62 & 0.54 & 5.82 & (0.37) \\
$1 < M < 3$ & 1.94 & 0.49 & 6.12 & (0.31) \\
\hline\hline
\end{tabular}
\end{center}
\label{table:table1}
\end{table}

The second-order bias factors $\chi_{10}$ and $\chi_{01}$ of the dark matter haloes at
$z=0$ can be readily computed from Eq.(\ref{eq:chiestimate}) using the best-fit values 
of $\la 3\eta^2|{\rm halo}\ra$ and $\la 5\zeta^2|{\rm halo}\ra$. The results are shown 
in Fig.\ref{fig:chihb} as the data points. Error bars indicate the scatter among the 
various realizations and, therefore, likely strongly underestimate the true uncertainty. 
The dashed curves indicate the predictions of the ESP formalism. The measurements, 
albeit of the same magnitude as the theoretical predictions, quite disagree with 
expectations based on our ESP approach, especially $\chi_{01}$ which reverses sign as 
the halo mass drops below $10^{14}\hmsun$.

\subsection{Interpretation of the measurements}

To begin with, we note that, if haloes were forming out of randomly distributed patches 
in the initial conditions, then both $\chi_{10}$ and $\chi_{01}$ would be zero since 
$\la 3\eta^2\ra= 3$ and $\la 5\zeta^2\ra = 5$ for random field points.

\begin{figure}
\centering
\resizebox{0.45\textwidth}{!}{\includegraphics{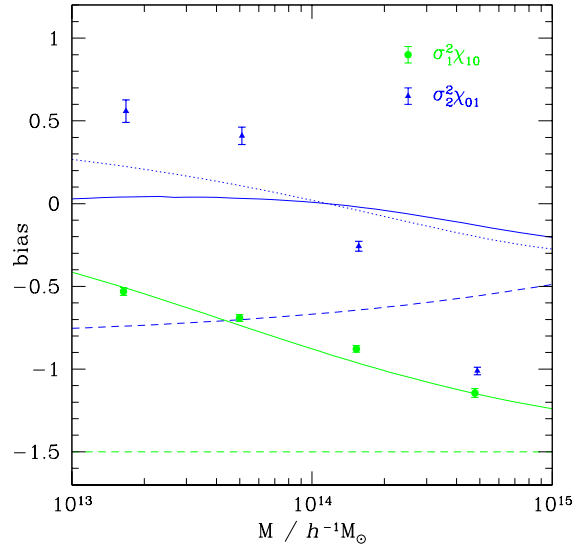}}
\caption{The bias factors $\sigma_1^2\chi_{10}$ and $\sigma_2^2\chi_{01}$ of dark 
matter haloes identified in the N-body simulations at $z=0$ are shown as filled
(green) circle and (blue) triangle, respectively. Error bars indicate the scatter 
among 6 realizations. The horizontal dashed (green) line at $-3/2$ and the dashed
(blue) curve are the corresponding ESP predictions. The dotted (blue) curve is 
$\sigma_2^2\chi_{01}$ in a model where halos are allowed to collapse in 
filamentary-like structures. The solid curves are our final predictions, which 
take into account the offset between peak position and proto-halo center-of-mass 
(see text for details).}
\label{fig:chihb}
\end{figure}

The measured dimensionless bias factor $\sigma_1^2\chi_{10}$ is always negative, which
indicates that halos collapse out of regions which have values of $\eta^2$ smaller than 
average. In our ESP approach, we assume that the center-of-mass position of proto-haloes 
exactly coincides with that of a local density peak, so that 
$\sigma_1^2\chi_{10}\equiv -3/2$. However, simulations indicate that, while there is a
good correspondence between proto-haloes and linear density peaks, the center-of-mass of 
the former is somewhat offset relative to the peak position
\citep[see e.g.][]{porciani/dekel/hoffman:2002a,ludlow/porciani:2011}. 
To model this effect, we note that, if the proto-halo is at a distance $R$ from a 
peak, then the average value of $3\eta^2$ is 
$\bigl\la3\eta^2\bigr\ra(R) = \epsilon_1^2(R) (\bigl\la3\eta^2|{\rm pk}\bigr\ra-3)$
(in analogy with the fact that the average density at a distance $R$ from a position 
where $\delta\equiv\delta_c$ is $\bigl\la\delta\bigr\ra(R)=\xi_\delta(R)\,\delta_c$). 
Assuming that the offset $R$ follows a Gaussian distribution, the halo bias factor is
\begin{equation}
\sigma_{1s}^2\chi_{10} = -\frac{3}{2}\sqrt{\frac{2}{\pi}}\int_0^\infty\!
\frac{dR}{\sigma}\left(\frac{R}{\sigma}\right)^2 e^{-R^2/2\sigma^2}\epsilon_1^2(R)\;.
\end{equation}
The rms variance $\sigma(M)$ of the offset distribution, which generally depends on the 
halo mass, can be constrained from our measurements of $\chi_{10}$ for dark matter haloes. 
The best-fit powerlaw function,
\begin{equation}
\sigma(M) = 2.50 \left(\frac{M}{10^{13}\hmsun}\right)^{0.063}\hmpc \;,
\end{equation}
turns out to be a weak function of halo mass. In unit of the (tophat) Lagrangian halo
radius, this translates into $\sigma/R_T\approx 0.79$ and $\approx 0.36$ for a halo mass 
$M=10^{13}$ and $10^{14}\hmsun$, respectively. 
The resulting theoretical prediction is shown as the solid curve in Fig.\ref{fig:chihb}
and agrees reasonably well with our data. This crude approximation demonstrates that an
offset between the proto-halo center-of-mass and the peak position can have a large
impact on the inferred value of $\chi_{10}$, since the latter is very sensitive to 
small-scale mass distribution.


Likewise, an offset between the proto-halo center-of-mass and the position of the linear 
density peak will also impact the measurement of $\chi_{01}$, yet cannot explain the 
observed sign reversal. In this regard, one should first remember that density peaks 
become increasingly spherical as $\nu\to\infty$. Nevertheless, while their mean ellipticity 
$\la e\ra$ and prolateness $\la p\ra$ converge towards zero in this limit, 
$\la v\ra =\la u e\ra$ approaches $1/5$ at fixed $u$ 
\citep[see Eq.(7.7) of][]{bardeen/bond/etal:1986}. Hence, $\la\zeta^2\ra=\la 3v^2+w^2\ra$ 
does not tend towards zero but rather unity, like for random field points. 
Consequently, $\sigma_2^2\chi_{01}\to 0$ in the limit $\nu\to\infty$. 
Secondly, at any finite $\nu$, our ESP approach predicts that $\chi_{01}$ be negative 
because we have assumed that proto-haloes only form near a density peak ($\lambda_3>0$, 
where $\lambda_1 \geq \lambda_2 \geq  \lambda_3 $ are the eigenvalues of 
$- \partial_i \partial_j \delta$). 
However, N-body simulations strongly suggest that a fraction of the proto-haloes collapse 
along the ridges or filaments connecting two density maxima, and that this fraction 
increases with decreasing halo mass \cite{ludlow/porciani:2011}. To qualitatively assess
the impact of such primeval configurations on $\chi_{01}$, we extend the integration domain 
in the plane $(v,w)$ to include all the points with $\lambda_2>0$ and $\lambda_3<0$ (but 
still require that the curvature $u$ be positive). This way we not only consider density 
peaks, but also extrema that correspond to filamentary configurations. The resulting  
curvature function $f(u,\alpha)$ can be cast into the compact form
\begin{align}
\label{eq:fua}
f(u,\alpha) &= \frac{1}{\alpha^4}
\Biggl\{\frac{e^{-{5\alpha u^2\over 2}}}{\sqrt{10\pi}}
\left(\alpha u^2-\frac{16}{5}\right) \\ 
&\quad +\frac{e^{-{5\alpha u^2\over 8}}}
{\sqrt{10\pi}}\left(31\alpha u^2+\frac{32}{5}\right)
+\frac{\sqrt{\alpha}}{2}\left(\alpha u^3-3u\right) \nonumber \\
& \qquad \times
\left[{\rm Erf}\left(\sqrt{\frac{5\alpha}{2}}\frac{u}{2}\right)+
{\rm Erf}\left(\sqrt{\frac{5\alpha}{2}}u\right)-1\right]\Biggr\} 
\nonumber \;,
\end{align}
The dotted curve in Fig.\ref{fig:chihb} shows $\sigma_2^2\chi_{01}$ when the filamentary 
configurations are included. While it agrees with the original ESP prediction at large 
halo mass, it reverses sign around $10^{14}\hmsun$ because, as the peak height decreases, 
configurations with $\lambda_3<0$ or, equivalently, large values of $\zeta^2$ become more 
probable. The solid curve takes into account, in addition to filamentary configurations,
an offset between the proto-halo and the peak position according to the simple prescription
discussed above. This is our final prediction for $\sigma_2^2\chi_{01}$. It is clearly at odds
with the measurements, which strongly suggest that $\sigma_2^2\chi_{01}$ can be very different
from zero for $M\gtrsim 10^{13}\hmsun$.


It is beyond the scope of this paper to work out a detailed description of the measurements.
Using a value of $R_G$ different than that obtained through the condition 
$\la\delta_G|\delta_T\ra = \delta_T$ has a large impact on the mass function, suggesting
that it will be difficult to get a good fit of both the mass function and the bias factors
$\chi_{10}$ and $\chi_{01}$.
Before concluding however, we note that, if the Lagrangian clustering of haloes also depends 
on $s_2(\vx)=s_{ij}(\vx)s^{ij}(\vx)$, where (in suitable units)
\begin{equation}
s_{ij}(\vx) = \partial_i\partial_j\phi(\vx) - \frac{1}{3}\delta_{ij}\delta(\vx) \;,
\end{equation}
then we are not measuring $\chi_{01}$ but some weighted and scale-dependent combination of 
both $\chi_{01}$ and the Lagrangian bias $\gamma_2$ associated with $s_2(\vx)$. Recent 
numerical work indeed suggests that $\gamma_2$ might be non-zero for massive haloes 
\citep{baldauf/seljak/etal:2012,chan/scoccimarro/sheth:2012,sheth/chan/scoccimarro:2013}. 
In this regards, our approach can furnish a useful cross-check of these results since it 
can provide a measurement of $\gamma_2$ which is independent of the bispectrum.

\section{Conclusion}
\label{sec:conclusion}

Dark matter haloes and galaxies are inherently biased relative to the mass density field, 
and this bias can manifest itself not only in $n$-point statistics such as the power spectrum
or bispectrum, but also in simpler one-point statistics. In this work, we took advantage of
this to ascertain the importance of certain nonlocal Lagrangian bias factors independently of 
a 2-point measurement. 
We extended the cross-correlation technique of \cite{musso/paranjape/sheth:2012} to
$\chi^2$-distributed variables, focusing on the quadratic terms $\eta^2(\vx)$ and $\zeta^2(\vx)$ 
(see Eq.\ref{eq:chivars}) which arise from the peak constraint and for which we have theoretical
predictions. In principle however, our approach could be applied to measure the Lagrangian bias 
factor associated with any $\chi^2$-distributed variable such as the tidal shear for instance.
We validated our method with peaks of Gaussian random field before applying it to a catalogue
of dark matter haloes with mass $M>10^{13}\hmsun$. 
Including an offset between the proto-halo center-of-mass and the peak position in the
modelling \citep[motivated by the analysis of][]{ludlow/porciani:2011}, we were able to reproduce 
our measurements of the nonlocal bias $\sigma_1^2\chi_{10}$. 
Our result $\chi_{10} < 0$ is consistent with the findings of \cite{ludlow/porciani:2011}, who 
demonstrated that proto-haloes with $M>10^{13}\hmsun$ preferentially form near initial density 
peaks ($\chi_{10}\equiv 0$ for a random distribution). 
However, we were unable to explain the measurements of $\sigma_2^2\chi_{01}$, even with the 
additional assumption that a fraction of the haloes collapse from filamentary-like structures 
rather than density peaks. We speculate that a dependence of the halo Lagrangian bias on 
$s_2({\bf x})$ might be needed to explain this discrepancy.

The dependence on $\eta^2({\bf x})$ induces a correction $-2\chi_{10}(\vk_1\cdot\vk_2)$ 
to the halo bias which, for collinear wavevectors $\vk_1$ and $\vk_2$ of wavenumber $0.1\hmpc$, 
is $\Delta b\approx 0.02$ (0.05) and $\approx 0.30$ (0.88) for haloes of mass $M=10^{13}$ and 
$10^{14}\hmsun$ at redshift $z=0$ (z=1), respectively. Relative to the evolved, linear halo 
bias $b_1^{\rm E}\equiv 1 + b_{100}$, the fractional correction is 
$\Delta b/b_1^{\rm E}\sim $2\% and $\sim 15$\% for the same low and high halo mass in the 
redshift range $0<z<1$. Hence, this correction can safely be ignored for $M=10^{13}\hmpc$, but 
it could become relevant at larger halo masses.


We also refined the ESP approach of PSD so that clustering statistics can be straightforwardly 
computed from the (effective) bias expansion Eq.(\ref{eq:dpk}) 
\citep[following the prescription detailed in][]{desjacques:2013}. 
We checked that the predicted halo mass function, from which all the bias factors can be derived, 
agrees well with the numerical data.
However, some of the model ingredients, especially the filtering of the density field, will 
have to be better understood if one wants to make predictions that are also accurate at small 
scales.

\section*{Acknowledgment}

V.D. would like to thank the Perimeter Institute for Theoretical Physics and CCPP at
New York University for their hospitality while some of this work was being completed 
there. M.B., K.C.C. and V.D. acknowledge support by the Swiss National Science 
Foundation.

\bibliography{references}

\appendix

\section{The curvature function of density peaks}
\label{app:cfunc}

The curvature function of density peaks is \citep{bardeen/bond/etal:1986}
\begin{align}
f(u,\alpha) &= \frac{1}{\alpha^4}
\Biggl\{\frac{e^{-{5\alpha u^2\over 2}}}{\sqrt{10\pi}}
\left(\alpha u^2-\frac{16}{5}\right) \\ 
&\quad +\frac{e^{-{5\alpha u^2\over 8}}}
{\sqrt{10\pi}}\left(\frac{31}{2}\alpha u^2+\frac{16}{5}\right)
+\frac{\sqrt{\alpha}}{2}\left(\alpha u^3-3u\right) \nonumber \\
& \qquad \times
\left[{\rm Erf}\left(\sqrt{\frac{5\alpha}{2}}\frac{u}{2}\right)+
{\rm Erf}\left(\sqrt{\frac{5\alpha}{2}}u\right)\right]\Biggr\} 
\nonumber \;.
\end{align}
Note that \cite{desjacques/crocce/etal:2010} introduced the extra variable $\alpha$
in order to get a closed form expression for their 2-point peak correlation, while 
\cite{desjacques:2013} showed that $\alpha\neq 1$ can be interpreted as a long-wavelength 
perturbation in $\zeta^2(\vx)$. 

\begin{figure}
\centering
\resizebox{0.45\textwidth}{!}{\includegraphics{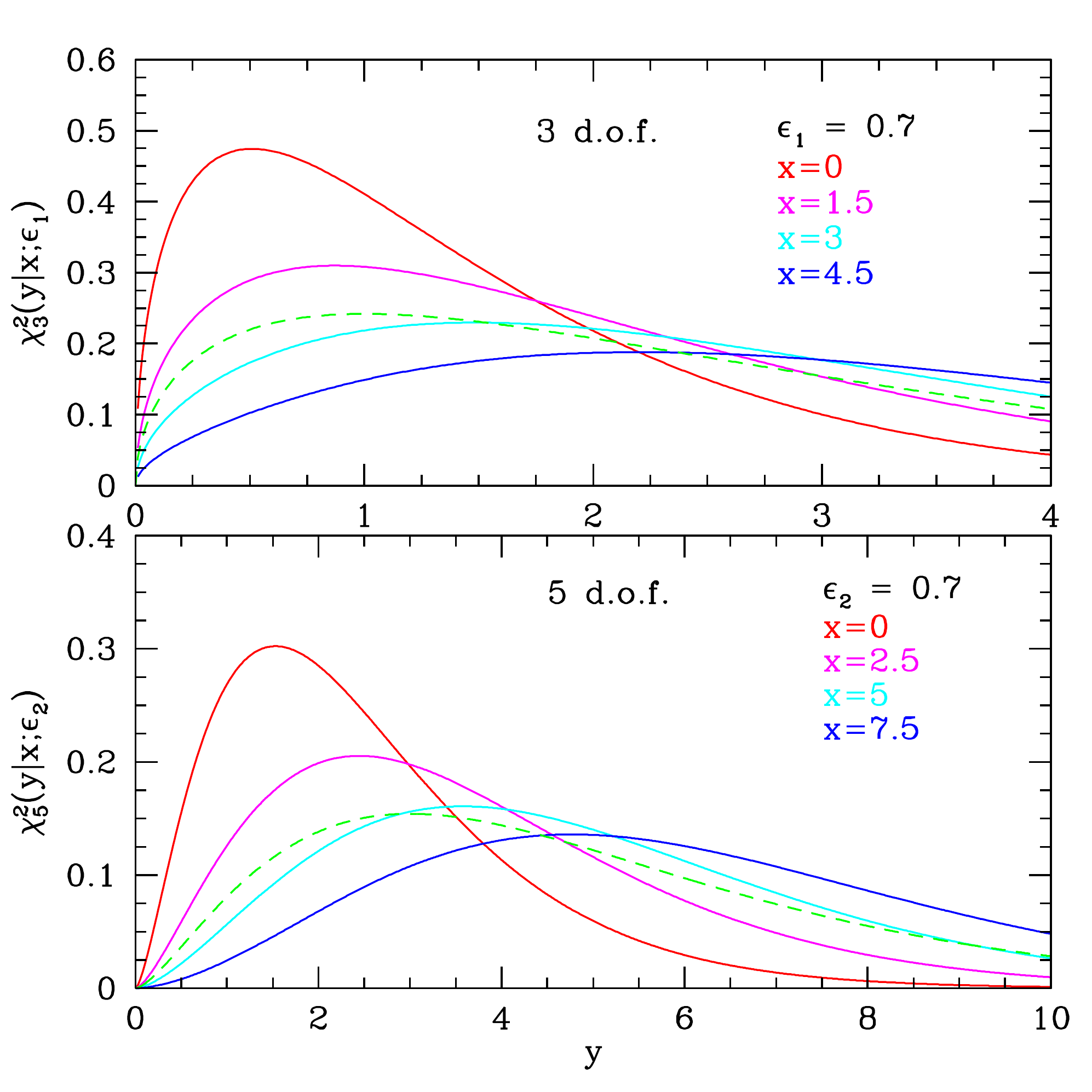}}
\caption{Conditional chi-squared distribution $\chi_k^2(y|x;\epsilon)$ for 
3 and 5 degrees of freedom. Results are shown for several values of $x$ 
and a fixed cross-correlation coefficient $\epsilon=0.7$. The dashed (green) 
curve represents the unconditional distribution $\chi_k^2(y)$.}
\label{fig:chixy}
\end{figure}

\section{Bivariate $\chi^2$ distributions}
\label{app:chidist}

We take the following expression for the bivariate $\chi^2$-distribution 
\citep{gunst/webster:1973}
\begin{align}
\chi^2_k(x,y;\epsilon) &= 
\frac{(xy)^{k/2-1}}{2^k\Gamma^2(k/2)}\left(1-\epsilon^2\right)^{-k/2}
e^{-\frac{x+y}{2(1-\epsilon^2)}} \\
&\qquad \times {}_0F_1\!\left(\frac{k}{2};\frac{\epsilon^2 xy}{4(1-\epsilon^2)^2}\right)
\nonumber \;,
\end{align}
where $x$ and $y$ are distributed as $\chi^2$-variables with $k$ d.o.f., 
$\epsilon^2\leq 1$ is their correlation and ${}_0F_1$ is a confluent
hypergeometric function. On using the fact that modified Bessel functions of 
the first kind can be written as $I_\alpha(x)=i^{-\alpha}J_\alpha(ix)$, where
\begin{equation}
J_\alpha(x) = \frac{(x/2)^\alpha}{\Gamma(\alpha+1)}\, 
{}_0F_1(\alpha+1;-\frac{x^2}{4})\;,
\end{equation}
the bivariate $\chi^2$-distribution can be reorganized into the product
\begin{equation}
\chi^2_k(x,y;\epsilon) = \chi^2_k(x) \chi^2_k(y|x;\epsilon) \;,
\end{equation}
where
\begin{equation}
\chi^2_k(y|x;\epsilon)= 
\frac{e^{-\frac{y+\epsilon^2 x}{2(1-\epsilon^2)}}}{2(1-\epsilon^2)}
\left(\frac{y}{\epsilon^2 x}\right)^{\alpha/2} 
I_\alpha\!\left(\frac{\epsilon\sqrt{x y}}{1-\epsilon^2}\right) \;,
\label{eq:chixy1}
\end{equation}
and $\alpha=k/2-1$. This conditional distribution takes a form similar to 
that of a non-central $\chi^2$-distribution $\chi^{2'}_k(x;\lambda)$, where 
$\lambda$ is the non-centrality parameter. Fig.\ref{fig:chixy} displays 
$\chi_k^2(y|x;\epsilon)$ for several values of $x$, assuming $k=3$ and 5. 
Note that $\chi_k^2(y|x=k;\epsilon)$ is different from $\chi_k^2(y)$.

Using the series expansion of $\chi^{2'}_k(x;\lambda)$ in terms of Laguerre 
polynomials \citep{tiku:1965}, we arrive at
\begin{align}
\chi_k^2(y|x;\epsilon) &= 
\frac{e^{-\frac{y}{2(1-\epsilon^2)}}}{2(1-\epsilon^2)^{\alpha+1}}
\left(\frac{y}{2}\right)^\alpha \\
& \qquad \times
\sum_{j=0}^\infty \frac{\left(\frac{-\epsilon^2}{1-\epsilon^2}\right)^j}
{\Gamma\left(\frac{1}{2}k+j\right)} \left(\frac{x}{2}\right)^j
L_j^{(\alpha)}\!\!\left[\frac{y}{2(1-\epsilon^2)}\right] \nonumber \;.
\label{eq:chixy2}
\end{align}
This series expansion is used to obtain the right-hand side of Eq.(\ref{eq:seriesL}).

\label{lastpage}

\end{document}